\documentclass[floatfix,aip,amsmath,amssymb,reprint,graphicx]{revtex4-2}

\usepackage{lineno}
\usepackage{booktabs}
\usepackage{multirow}
\usepackage{siunitx}
\def\ET{$\alpha$-(BEDT-TTF)$_2$I$_3$}
\def\cm{cm$^{-1}$}
\usepackage{graphicx}%
\usepackage{color}
\usepackage{epstopdf}
\usepackage{amssymb}
\usepackage{amsmath}
\usepackage{amsfonts}
\usepackage{xcolor}

\usepackage{color}
\usepackage[colorlinks,bookmarks=false,citecolor=darkblue,linkcolor=red,urlcolor=blue]{hyperref} 
\definecolor{darkred}{rgb}{0.7,0.0,0.0}

\definecolor{darkblue}{rgb}{0,0.02,0.45}

\definecolor{darkgreen}{rgb}{0.02,0.45,0.0}

\definecolor{violet}{rgb}{0.8,0.2,0.6}

\begin{document}

\title{Temperature-dependent generalized ellipsometry of the metal-insulator phase transition in low-symmetry charge-transfer salts} 

\author{Achyut Tiwari}
\author{Bruno Gompf}
\author{Martin Dressel}
\email{dressel@pi1.physik.uni-stuttgart.de}
\affiliation{1. Physikalisches Institut, Universit{\"a}t Stuttgart, Pfaffenwaldring 57, 70569 Stuttgart, Germany}

\begin{abstract}
Determining the optical and electronic properties of strongly anisotropic materials with symmetries below orthorhombic remains challenging; generalized ellipsometry is a powerful technique in this regard. Here, we employ Mueller matrix spectroscopic and temperature-dependent ellipsometry to determine the frequency dependence of six components of the dielectric-function tensor of the two-dimensional charge-transfer salt $\alpha$-(BEDT-TTF)$_2$I$_3$ across its metal-insulator transition. Our results offer valuable insights into temperature-dependent changes of the components of the spectroscopic dielectric-function tensor across the metal-insulator transition. This advanced method allows extension to other electronic transitions.

\end{abstract}


\maketitle 


The two-dimensional organic conductor \ET\ has been subject of keen interest among solid-state physicists for several decades due to its rich variety of electronic properties \cite{Dressel2020}. These properties include a correlation-driven metal-insulator transition \cite{takahashi2006charge}, superconductivity \cite{tajima2002effects}, photo-induced phase transitions \cite{iwai2007photoinduced}, and zero-gap semiconductivity \cite{mori2009requirements} with massless Dirac-like fermions \cite{Peterseim, Tajima_2007, Tajima}. The material's intriguing behavior has spurred extensive research aimed at understanding the underlying mechanisms governing its electronic responses. \ET\ is the prime example among two-dimensional organic conductors, which exhibits
a metal-insulator-transition at $T_\mathrm{CO}$~=~135 K \cite{takahashi2006charge}. This electronically driven metal-insulator phase transition involves an abrupt change from a metallic state to an insulating state \cite{Yue10} without any remarkable crystallographic transition \cite{kakiuchi2007charge, Thomas86}. Despite significant progress in understanding the metal-insulator transition in \ET\ over the last years, the underlying mechanism remains poorly understood on a microscopic scale.

In this regard, spectroscopic ellipsometry has been proven to be an excellent tool to gain deeper insights into microscopic characteristics \cite{Hoevel2010, Voloshenko2018, Voloshenko2019}. However, understanding the microscopic properties using macroscopic measurement becomes rather challenging in \ET\ due to the low symmetry of the triclinic crystals. In crystals with symmetry lower than orthorhombic, the optical response from various directions tends to mix up, which needs proper analysis to accurately disentangle the contributions from different directions. Generalized ellipsometry has been used to determine the dielectric function of low-symmetry crystals such as $\alpha$-PTCDA \cite{ALONSO200223}, pentacene \cite{dressel2008kramers}, BiFeO3 \cite{BFO}, CdWO$_4$ \cite{CDW}, Ga$_2$O$_3$ \cite{MS2016} and K$_2$Cr$_2$O$_7$ \cite{HOFER2014111,Strum2020}.

In this letter, we present a comprehensive temperature-dependent study employing generalized ellipsometry to investigate the anisotropic dielectric properties and the charge-order phase transition in \ET. 
In the first step, Mueller matrix (MM) ellipsometry is conducted at room temperature to appropriately address the anisotropic optical behavior of \ET. It allows us to determine the crystal orientation relative to the laboratory frame and extract the real and imaginary parts of the six components of the dielectric-function tensor, including the unit-cell angles. Subsequently, we analyze the temperature-dependent changes in the dielectric functions by applying spectroscopic ellipsometry, revealing the metal-insulator phase transition. The crystal orientation acquired via MM is used in this analysis, since the phase transition is solely driven by effective electronic correlation without breaking crystallographic symmetry. Based on the derived temperature-dependent dielectric functions, we discuss the changes appearing in the various excitations across the metal-insulator transition.

\begin{figure}
	\centering
	\includegraphics[width=0.85\columnwidth]{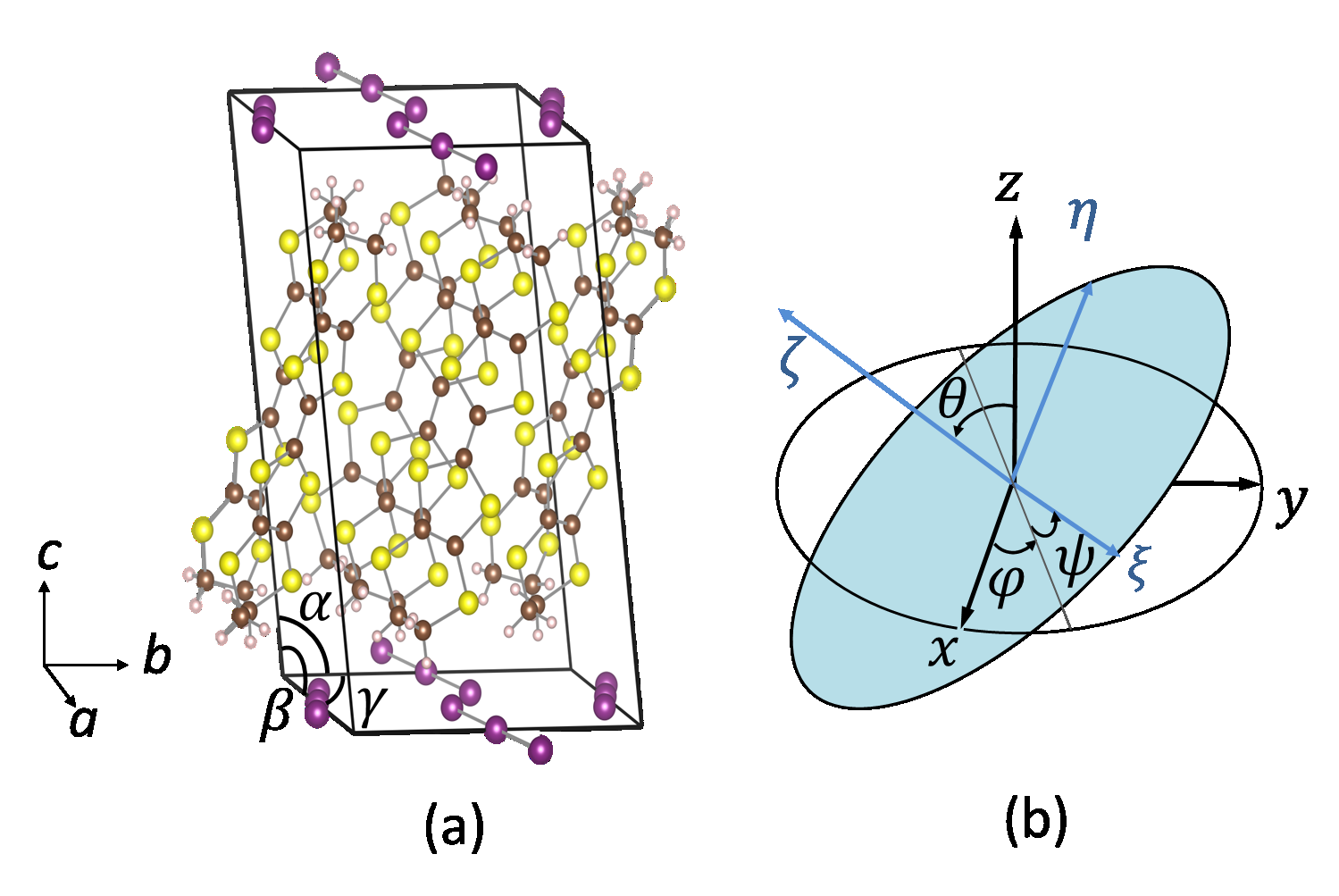}
	\caption{(a) Unit cell of \ET. $\alpha$, $\beta$ and $\gamma$ are the triclinic unit-cell angles, and crystallographic axes are indicated as $a$, $b$ and $c$ \cite{kobayashi1984crystal, kakiuchi2007charge}. (b) Schematic representation of the orientation of orthogonal auxiliary frame with respect to laboratory frame (\textit{x},\textit{y},\textit{z}), defined by Euler angles $\phi$, $\theta$ and $\psi$.}	
	 \label{fig: Crystal structure and coordinate transformation}
\end{figure}

The crystal structure of \ET\ is triclinic, with alternating conducting donor layers of BEDT-TTF molecules and insulating anions (I$_\mathrm{3}^-$) along the $c$-direction \cite{bender1984bedt, kobayashi1984crystal, kakiuchi2007charge}, as depicted in Fig. \ref{fig: Crystal structure and coordinate transformation}(a). Due to the low symmetry, the electrical and optical properties are characterized by a small in-plane and a substantial out-of-plane anisotropy \cite{Ivek2011}. For crystals with symmetry lower than orthorhombic, the non-orthogonal tilting of the crystallographic axes leads to a non-diagonalizable dielectric tensor of six compenents for non-magnetic material.
In this case, the complex dielectric function is given by the frequency-dependent $3 \times 3$ tensor \cite{born2013principles}:
\begin{equation}
\textbf{$\tilde{\varepsilon}$} =
\begin{pmatrix}
\tilde{\varepsilon}_{11} & \tilde{\varepsilon}_{12} & \tilde{\varepsilon}_{13} \\
\tilde{\varepsilon}_{12} & \tilde{\varepsilon}_{22} &\tilde{\varepsilon}_{23} \\
\tilde{\varepsilon}_{13} & \tilde{\varepsilon}_{23} & \tilde{\varepsilon}_{33}
\end{pmatrix}  .
\end{equation}
For these crystals, the optical response depends on the direction of the incident light and the crystal orientation, which necessitates careful treatment to accurately disentangle the different contributions from each direction \cite{Jellison:22, dressel2008kramers}.

Ellipsometry is a powerful technique to explore the optical properties of isotropic samples and thin films \cite{drude1887ueber, aspnes1997accurate, fujiwara2007spectroscopic}. Generalized ellipsometry is a robust tool to investigate anisotropic materials, including monoclinic and triclinic crystals \cite{azzam1974application, schubert1996extension, schubert1998generalized, schubert2002generalized, dressel2008kramers, schmidt2009monoclinic, Strum2020}. It describes the  interaction of electromagnetic waves with anisotropic samples within the Jones- or Mueller-matrix formalism. Here, we use the Stokes vector formalism, which represents the connection between the real-valued $4 \times 1$ Stokes vectors before and after the interaction with the sample using the $4 \times 4$ real-valued Mueller matrix \cite{born2013principles, tompkins2005handbook, azzam1995ellipsometry}. The Stokes vector is defined by the experimentally accessible intensities as
\begin{equation}
\textbf{S} = \left( \begin{array}{c}
S_\mathrm{0} \\
S_\mathrm{1} \\
S_\mathrm{2}\\
S_\mathrm{3}\\
\end{array} \right) = \left( \begin{array}{c}
I_\mathrm{p} + I_\mathrm{s} \\
I_\mathrm{p} - I_\mathrm{s} \\
I_\mathrm{+45^\circ} - I_\mathrm{-45^\circ} \\
I_\mathrm{CR} - I_\mathrm{CL} \\
\end{array} \right) \quad .
\end{equation}
Where $I_\mathrm{p},I_\mathrm{s}, I_\mathrm{+45^\circ}, I_\mathrm{-45^\circ}, I_\mathrm{CR}, I_\mathrm{CL} $ are the intensities of  p, s, $+45^\circ$, $-45^\circ$, and right-, and left-handed circularly polarized light, respectively \cite{fujiwara2007spectroscopic}. The  $4 \times 4$ Mueller matrix \textbf{M} transforms the incident vector $\textbf{S}_\mathrm{in}$ to the outgoing vector $\textbf{S}_\mathrm{out}$ according to
$
\textbf{S}_\mathrm{out} = \textbf{M}\; \textbf{S}_\mathrm{in}
$.
The Mueller matrix elements contain the entire information of the optical response, making it a valuable method for characterizing the complex optical behavior of \ET.
\begin{figure*} [ht]
	\centering
	\includegraphics[width=1\linewidth]{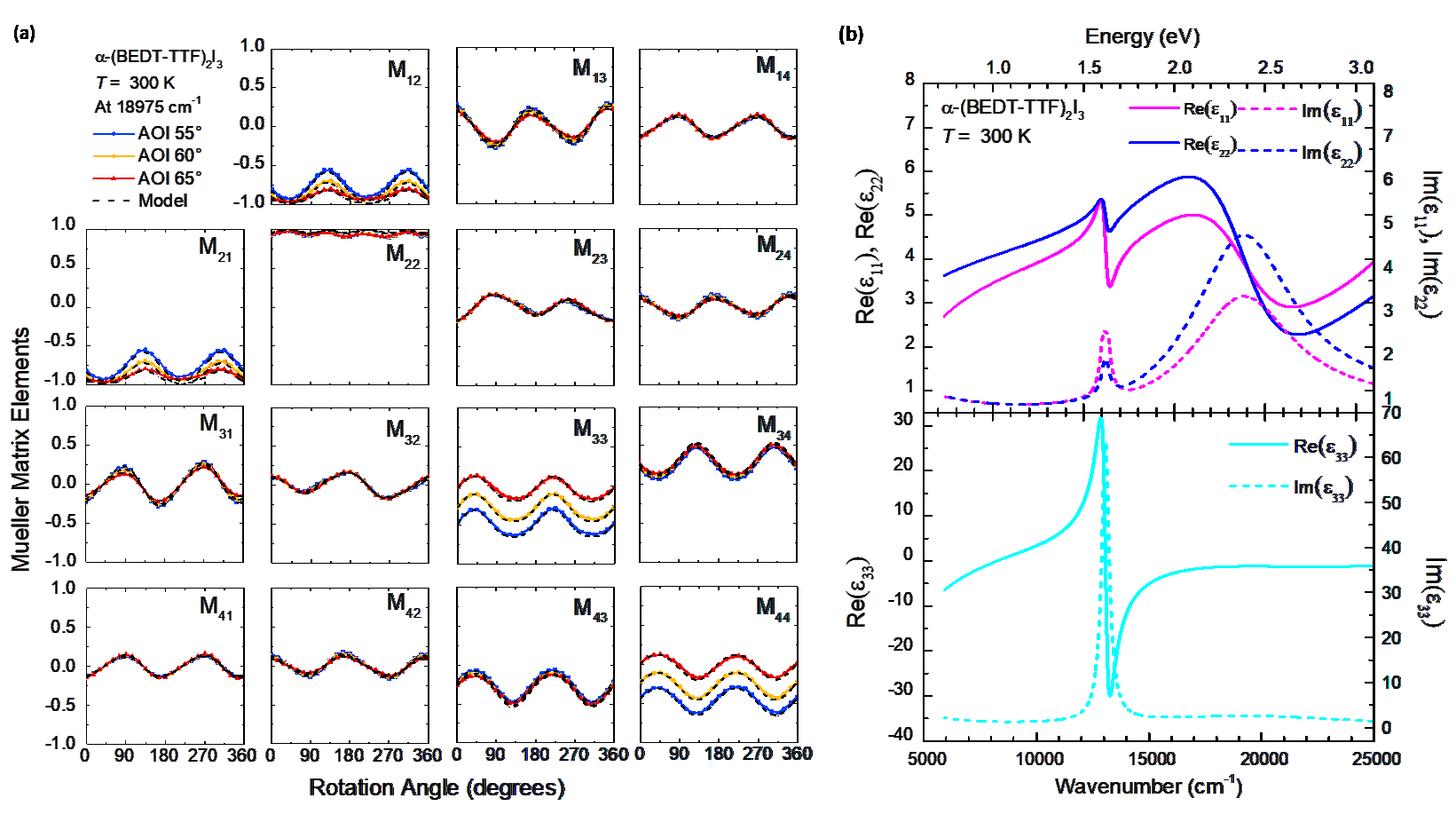}
	\caption{(a) Experimental (colored solid lines and dots) and best match calculated (black dashed lines) Mueller matrix elements for \ET\ versus sample azimuth rotation at 2.35 eV for three different angles of incidence ($55^\circ$, $60^\circ$, and $65^\circ$). The sample was rotated azimuthally from $0^\circ$ to $360^\circ$ in steps of $15^\circ$. Note all the Mueller matrix elements are normalized to M$_\mathrm{11}$. (b) Real (solid lines, left axis) and imaginary (dashed lines, right axis) parts of the dielectric-function tensor $\varepsilon_\mathrm{11}$, $\varepsilon_\mathrm{22}$ (upper panel), $\varepsilon_\mathrm{33}$ (lower panel) of \ET. Due to the availability of only one crystal face for measurements, the uncertainty in the out-of-plane components is greater than that of the in-plane components. Nevertheless, the level of sensitivity is adequate to support the conclusion (see supplement document Table-S1).}
	\label{fig:Mueller matrix elements}
\end{figure*}

The investigated \ET\ single crystals were grown via standard electrochemical methods, described in Ref. \onlinecite{bender1984synthesis}. The compound crystallizes in the space group P$\bar{1}$ \cite{kakiuchi2007charge}. Single crystals are shaped like plates, measuring 3 $\times$ 2 mm in lateral dimensions and with a thickness 67 $\mu$m. Before measurements, the samples were visually inspected under an optical microscope to ensure they were clean and free from defects.
Mueller matrix ellipsometry measurements were performed on a state-of-the-art dual rotating compensator ellipsometer (J. A. Woollam RC2), equipped with micro-focus lenses. The experiments with this ellipsometer were carried out at room temperature in the range 300 nm - 1690 nm. The angles of incidence were $55^\circ$, $60^\circ$, and $65^\circ$, while the sample was rotated azimuthally from $0^\circ$ to $360^\circ$ in steps of $15^\circ$. This comprehensive approach allowed for a thorough characterization of the anisotropic optical properties including the crystallographic orientation.

In order to investigate the dielectric-function tensor of optically anisotropic material, the oriented dipole approach and a more recent general approach that involves characterizing the dielectric function through the distribution of dipole interactions for each excitation has been employed \cite{MS2016, Strum2020}. Here, for a proper analysis, it is essential to distinguish three coordinate frames: Firstly, the measurements are executed in the laboratory coordinates (\textit{x},\textit{y},\textit{z}), established by the plane of incidence (\textit{x},\textit{z}) and the sample surface (\textit{x},\textit{y}). Secondly, the macroscopic optical response is characterized by the second-rank tensor of the complex dielectric function $\tilde{\varepsilon}$ (eq. 1). Lastly, any microscopic description relies on the electronic system, with axes described by unit vectors \textbf{\emph {x}}, \textbf{\emph {y}} and \textbf{\emph{z}}, the dielectric polarizability \textbf{P} along $\mathbf{\hat{e}} = \hat{e}_x \mathbf{x} + \hat{e}_y \mathbf{y} + \hat{e}_z \mathbf{z}$ is given by $\mathbf{P}_{\hat{e}} = \rho_{\hat{e}} (\hat{e} \mathbf{E}) \hat{e}\ $. The complex-valued polarizability functions $\textbf{$\rho$}_\mathrm{\hat{e}}$  may vary with photon energy and must be Kramers-Kronig consistent.
 The linear polarization response, with $n$ the number of excitations along their respective polarization directions, is given by the superposition:
     \begin{equation}
\mathbf{P} = \sum_{l=1}^{n} \mathbf{P}_{\hat{e}_l} = \sum_{l=1}^{n} \rho_{\hat{e}_l} (\hat{e}_l \otimes \hat{e}_l) \mathbf{E} = \chi \mathbf{E},
\end{equation}

In the laboratory frame, the macroscopic polarization \textbf{P} is related to the electric  field vector \textbf{E} by the second-rank dielectric tensor $\tilde{\varepsilon}$:
$
\textbf{P} =  (1- \tilde{\varepsilon})\textbf{E}= \textbf{$\chi$} \textbf{E},  $ where \textbf{$\chi$} is the second-rank susceptibility tensor. 
From the above equations, all the six components of the symmetric dielectric tensor can be deconvoluted. To obtain the microscopic electronic properties from measurements conducted in the laboratory frame, a transformation process is required. The Cartesian coordinate system has to  undergo a rotation by the Euler angles $\phi$, $\theta$ and $\psi$, transforming it into an auxiliary coordinate system ($\xi$, $\eta$, $\zeta$)
[Fig. \ref{fig: Crystal structure and coordinate transformation}(b)] \cite{SchubertIR, tompkins2005handbook}. For orthorhombic, tetragonal, hexagonal, and trigonal systems, $\phi$, $\theta$ and $\psi$ can be chosen in a manner that the tensor $\tilde{\varepsilon}$  is diagonal in ($\xi$, $\eta$, $\zeta$). In the case of monoclinic and triclinic systems, complex dielectric tensor $\tilde{\varepsilon}$ can not be diagonalize in general.

The six independent components of the dielectric tensor can be extracted by analyzing all ellipsometry data at the same wavelength from multiple azimuth angles and multiple angles of incidence for all energies simultaneously, using a wavelength-by-wavelength approach. Additionally, an independent set of Euler-angle parameters is utilized to describe the orientation of the crystal axes and the elements of the dielectric tensor.
In this manner, the spectroscopic dielectric function tensor for triclinic \ET, represented by the tensor $\tilde{\varepsilon}$, is characterized by six components, as shown in Equation 1. In this context, the indices (1,2,3) correspond to the chosen \ET\ system.

In the next step, spectroscopic ellipsometry measurements were performed using a rotating analyzer ellipsometer (J. A. Woollam VASE) equipped with a customized liquid-helium flow cryostat to investigate the temperature-dependent optical properties. The cryostat restricts the angle of incidence to $70^\circ$. Window effects were corrected by reference measurements on a silicon sample. Only the standard ellipsometry has been performed for temperature-dependent measurement as variable angles of incidence and azimuthal rotations were not feasible due technical complexities imposed by crysostat and sample. 
In our analysis, the crystal orientation and optical anisotropy obtained from the Mueller matrix  ellipsometry at room temperature 
could be utilized since no significant structural transition occurs across the metal-insulator transition
according to x-ray diffraction measurements \cite{kakiuchi2007charge, Thomas86}. 
For data analysis, the CompleteEase software by J. A. Woollam Co. Inc. was used.

\begin{figure*}[!]
    \centering
    \includegraphics[width=0.75\linewidth]{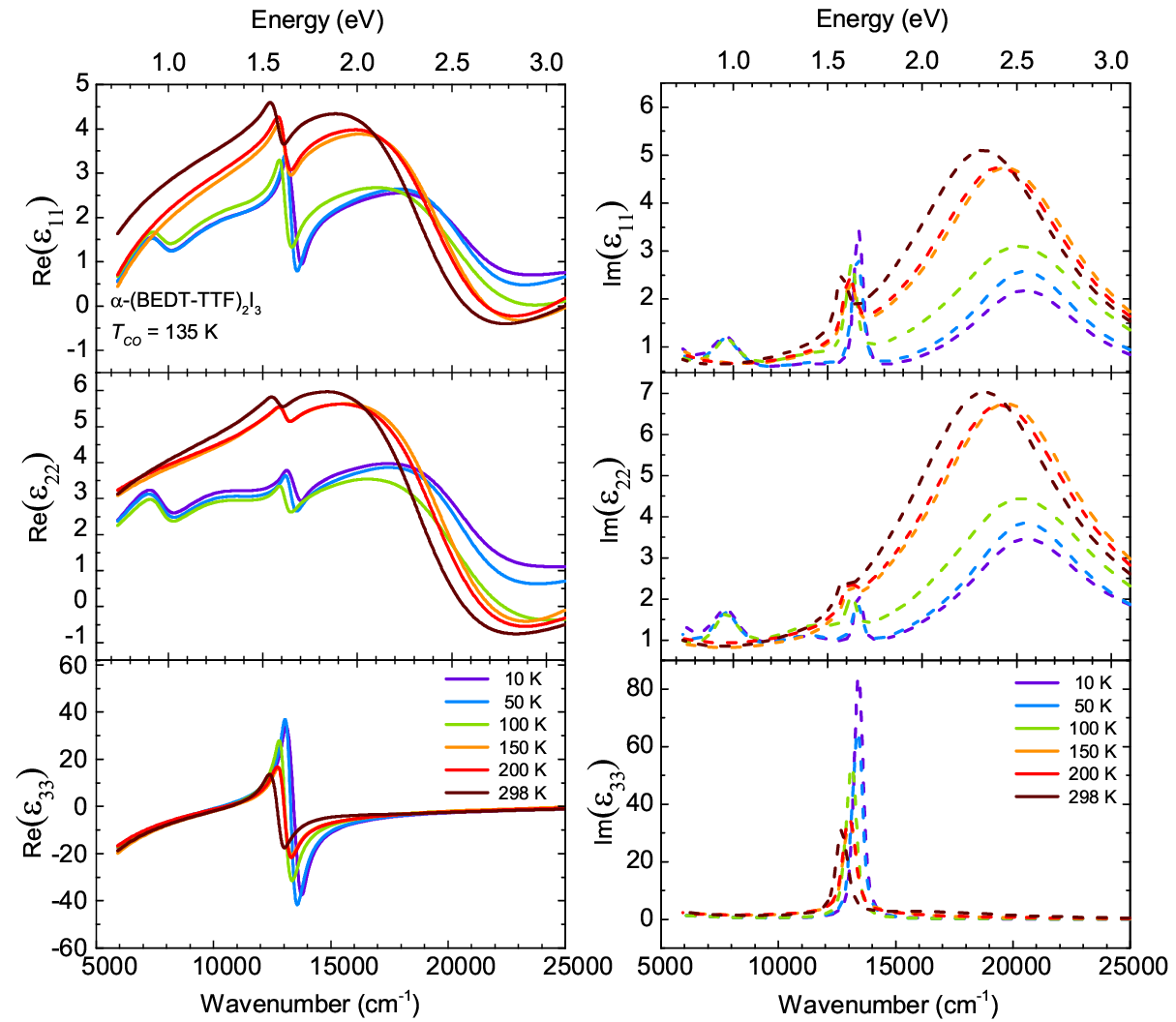}
    \caption{Real (left panels) and imaginary (right panels) parts of the dielectric functions $\varepsilon_\mathrm{11}$, $\varepsilon_\mathrm{22}$, $\varepsilon_\mathrm{33}$ of \ET\ vs frequency, for different temperatures across metal-insulator transition.}
    \label{fig:temp dependent dielectric field tensor}
\end{figure*}

In order to extract the dielectric functions, all Mueller matrix data were analyzed using a semi-infinite substrate model, representing the properties of \ET. The possibility of any surface effect can be neglected in the analysis, as the sample surface is perfectly smooth and does not effect the main results.
The crystallographic orientation relative to the laboratory frame was precisely determined by fitting the Mueller matrix elements obtained from various angles of incidence and azimuth rotations with a model dielectric function tensor.  This model incorporates set of Drude term, Tauc-Lorentz, and Lorentz oscillators, capturing the intricate anisotropic behavior of \ET. The Drude component accounts for the contribution of free charge carriers to the dielectric function in metallic states. Tauc-Lorentz and Lorentz oscillators have been employed to characterize intramolecular and intermolecular excitations, respectively, within the visible and near-infrared regions. Pole functions were incorporated to account for the dispersion in the real parts of each tensor element caused by absorption at higher energies beyond the measured spectral range. The oscillators give the Kramers-Kronig consistent dielectric functions. The experimental and simulated Mueller matrix elements exhibited excellent agreement, as depicted in Fig. \ref{fig:Mueller matrix elements}(a), validating the robustness of the fit. The Mueller matrix elements exhibit the approximate symmetry in the off-diagonal elements, corresponding to an antisymmetric Jones matrix. This behavior is generally not expected for triclinic systems but is typical for monoclinic systems \cite{ARTEAGA2014}. This observation is consistent with the fact that one of the crystal angles, $\gamma$, is close to $90^\circ$. The parameters of the model dielectric function are summarized in Table-S1. The crystal inclination with respect to the laboratory frame, characterized by the Euler angles, was found to be $\phi = 57.47^\circ \pm  0.27^\circ$, $\theta = 12.08^\circ \pm 0.17^\circ$ and $\psi = 84.04^\circ \pm 0.40^\circ$. The precise orientation is crucial for the analysis and interpretation of the anisotropic optical response of \ET.

Fig. \ref{fig:Mueller matrix elements}(b) displays the on-diagonal components of dielectric function tensor at room temperature, as obtained by the Mueller matrix measurements. Notably, in the \ET\  crystal system (a,b,c), on-diagonal components  $\varepsilon_\mathrm{11}$, $\varepsilon_\mathrm{22}$, approximately corresponds to in-plane (a,b) dielectric response, whereas $\varepsilon_\mathrm{33}$ reflects the out-of-plane (c) dielectric response. The real and imaginary parts of  $\varepsilon_\mathrm{11}$, $\varepsilon_\mathrm{22}$, $\varepsilon_\mathrm{33}$ revealed intriguing optical properties, exhibiting weak in-plane anisotropy and a pronounced out-of-plane anisotropy. In the \emph{a} and \emph{b}-directions, two absorption peaks are observed, with a broad prominent peak at 19,350 \cm\ and a smaller narrower peak at 13,000~\cm. The overall absorption in the visible spectrum is slightly higher in \emph{b}-direction and shifts towards lower frequencies when compared to the \emph{a}-direction. Conversely, the absorption peak at \text{13,000 \cm} is more pronounced in the \emph{a}-direction than in the \emph{b}-direction. The feature  at 19,350 \cm\ can be attributed to intra-molecular excitations, and the peak at 13,000 \cm\ to inter-molecular excitations. The results are in good agreement with Helberg's early study on electronic excitations in \ET,  who reports similar broad and pronounced absorption peaks in visible range and less pronounce absorption peaks along the \emph {a} and \emph{b}-directions \cite{Helberg1985, Helberg1987, Helberg1987BBG}. The strongly increased absorption peak at 13,000~\cm\  in \emph{c}-direction reflects the molecular stacking along that axis, leading to a stronger inter-molecular absorption.


As mentioned, low-temperature measurements were possible only at a single angle of incidence (70$^\circ$), therefore we are restricted to standard ellipsometric parameters, i.e., Psi and Delta were measured. In the analysis of the temperature-dependent ellipsometric data, different models were employed for the metallic and insulating states. In the metallic state, the analysis incorporates a Drude term, Tauc-Lorentz, and Lorentz oscillators in all three directions. For the insulating state, the Drude term is replaced by a Lorentz oscillator. The experimental data are in excellent agreement with the generated data (see supplement document Fig.-S2), demonstrating the accuracy and suitability of the chosen models. The mean-square error (MSE) and parameters of the model dielectric function are summarized in supplement document, (Table. S2-S7).

Although the crystallographic structure remains unaltered, the temperature-dependent dielectric functions exhibit pronounced changes across the charge-order transition at 135 K, as shown in Fig.~\ref{fig:temp dependent dielectric field tensor}. The observed variations are solely caused by effective electronic correlations. In the insulating state, the intra-molecular excitations are weaker and shift towards higher energies in both the  \emph {a}- and \emph {b}-directions. In contrast, the inter-molecular excitations in the \emph {a}-, \emph {b} directions are more intense and shift towards higher energies in the insulating state. This effect is even more pronounced in the \emph {c}-direction, pointing to significant modification in the material's electronic interactions in the stacking direction. While the shifts in the peak positions and the increase of intensities  is something to be expected at lower temperatures, the abrupt ermergence of a destinct new peak at approximately 7500~\cm\  in \emph {a}- and \emph {b}-directions in the insulating phase points towards something directly correlated to the metal-insulator transition. Previous results from Yakushi {\it et al.} suggested that this additional peak might arise from slight structural distortions across the phase transition \cite{yakushi1987temperature}. These small structural changes seem to be invisible by x-ray diffraction studies \cite{kakiuchi2007charge,NOGAMI1986367}. The structural modifications are attributed to alterations of the local anion-molecular interaction which occur when charge ordering sets in. Small displacements of anions lead to variations in the hydrogen bonding between anions and BEDT-TTF molecules, as elucidated by Alemany, Pouget and Canadell \cite{Alemany,Pouget18}. With the help of ellipsometry these slight structural modifications at the metal insulator transition can easily be analyzed. They only happen in the  \emph {ab}-plane, therefore no contribution in the \emph {c}-direction can be seen, the overall triclinic structure of the crystal remains unchanged.

The here presented comprehensive study on \ET\ demonstrates that temperature-dependent generalized ellipsometry is a powerful tool for characterizing metal-insulator transitions, even in strongly anisotropic materials.
We first employed Mueller matrix ellipsometry at room temperature for a deeper understanding of the optical response of the triclinic crystals. By a detailed analysis of the dielectric functions, we gained valuable insights into the nature of inter-molecular and intra-molecular excitations, observed in the near-infrared and visible spectral regions, respectively. The subsequent temperature-dependent investigations demonstate the high sensitivity of ellipsometry.
While continues shifts in the peak positions is something to be expected at lower temperatures, the abrupt emergence of a destinct new peak in \emph {a}- and \emph {b}-directions in the insulating phase points towards slight structural modifications not identified by x-ray diffraction.

We acknowledge the technical support by Gabriele Untereiner and thank Dieter Schweitzer for providing the crystals.
This work is supported by the Deutsche Forschungsgemeinschaft (DFG) under Grant No. DR228/63-1 and GO642/8-1.

\section*{SUPPLEMENTARY MATERIAL}
See supplementary materials for a list of parameters used to fit the Mueller matrix elements by Drude and various oscillator models. Also shown are  spectra of the experimental and calculated Mueller matrix elements at different angles of incidence and for various azimuth rotations in steps of $45^{\circ}$. The room-temperature off-diagonal elements of the dielectric tensor are plotted as a function of frequency. We present the temperature dependent standard ellipsometric parameters and list model parameters of the oscillators used to fit the temperature dependent standard ellipsometric data. Spectra of the off-diagonal elements of the complex dielectric tensor are plotted for different temperatures.

\section*{DATA AVAILABILITY}

 The data that support the findings of this study are available within the article and its supplementary material.

%

\end{document}